\documentclass[useAMS,usenatbib]{mn2e}

\usepackage[dvips]{graphicx}   
\usepackage{times}

\usepackage{epsfig}
\usepackage{amsmath,amssymb}
\usepackage{latexsym}
\usepackage{graphicx}
\usepackage{dcolumn}
\usepackage{bm}
\usepackage{epstopdf}
\def\reff@jnl#1{{\rm#1\/}}
\def\aj{\reff@jnl{AJ}}                  
\def\araa{\reff@jnl{ARA\&A}}            
\def\apj{\reff@jnl{ApJ}}                
\def\apjl{\reff@jnl{ApJ}}               
\def\apjs{\reff@jnl{ApJS}}              
\def\ao{\reff@jnl{Appl.Optics}}         
\def\apss{\reff@jnl{Ap\&SS}}            
\def\aap{\reff@jnl{A\&A}}               
\def\aapr{\reff@jnl{A\&A~Rev.}}         
\def\aaps{\reff@jnl{A\&AS}}             
\def\azh{\reff@jnl{AZh}}                
\def\baas{\reff@jnl{BAAS}}              
\def\jrasc{\reff@jnl{JRASC}}            
\def\memras{\reff@jnl{MmRAS}}           
\def\mnras{\reff@jnl{MNRAS}}            
\def\pra{\reff@jnl{Phys.Rev.A}}         
\def\prb{\reff@jnl{Phys.Rev.B}}         
\def\prc{\reff@jnl{Phys.Rev.C}}         
\def\prd{\reff@jnl{Phys.Rev.D}}         
\def\prl{\reff@jnl{Phys.Rev.Lett}}      
\def\pasp{\reff@jnl{PASP}}              
\def\pasj{\reff@jnl{PASJ}}              
\def\qjras{\reff@jnl{QJRAS}}            
\def\skytel{\reff@jnl{S\&T}}            
\def\solphys{\reff@jnl{Solar~Phys.}}    
\def\sovast{\reff@jnl{Soviet~Ast.}}     
\def\ssr{\reff@jnl{Space~Sci.Rev.}}     
\def\zap{\reff@jnl{ZAp}}                
\def\nat{\reff@jnl{Nature}}             
\newcommand{\be}{\begin{equation}}
\newcommand{\ee}{\end{equation}}
\newcommand{\bea}{\begin{eqnarray}}
\newcommand{\eea}{\end{eqnarray}}
\newcommand{\bi}{\begin{itemize}}
\newcommand{\ei}{\end{itemize}}

\def\G{G\mu/c^2}

\def\Om{\Omega_{\rm m}}
\def\Ol{\Omega_{\Lambda}}
\def\Gfrac{\frac{G\mu}{c^2}}
\def\zd{z_{\rm d}}
\def\zs{z_{\rm s}}
\def\Dd{D_{\rm d}}
\def\Ds{D_{\rm s}}
\def\Dds{D_{\rm ds}}

\title%
[Cosmic string lenses]%
{Direct Observation of Cosmic Strings\\ 
via their Strong Gravitational Lensing Effect: \\
I.~Predictions for High Resolution Imaging Surveys}
\label{firstpage}

\author%
[Gasparini et al]%
{Maria Alice Gasparini$^{1}$\thanks{E-mail:alice@physics.ucsb.edu}, 
Phil Marshall$^{1}$\thanks{E-mail:pjm@physics.ucsb.edu}, 
Tommaso Treu$^{1}$,
Eric Morganson$^{2}$ and
\newauthor Florian Dubath$^{3}$\\
$^{1}$Physics department, University of California, Santa Barbara, CA 93106, USA \\
$^{2}$KIPAC, P.O. Box 20450, MS29, Stanford, CA 94309, USA\\
$^{3}$KITP, University of California, Santa Barbara, CA 93106, USA}

\date{Accepted 2007 October 29. Received 2007 October 15; in original form 2007 July 12}

\pagerange{\pageref{firstpage}--\pageref{lastpage}}

\pubyear{2007}

\begin{document}

\maketitle


\begin{abstract}

We use current theoretical estimates for the density of long cosmic
strings to predict the number of strong gravitational lensing events
in astronomical imaging surveys as a function of angular resolution and survey 
area. We show that angular resolution is the most important
factor, and that interesting limits on the dimensionless string
tension $\G$ can be obtained by existing and planned surveys. At the
resolution of the Hubble Space Telescope ($0\farcs14$), it is
sufficient to survey of order a few square degrees -- well within reach of
the current HST archive -- to probe the regime $\G\sim10^{-7}$. If lensing
by cosmic strings is not detected, such a survey would improve the limit 
on the string tension by a
factor of two over that available from the cosmic microwave
background. Future high resolution imaging surveys, covering a few hundred
square degrees or more, either from
space in the optical or from large-format radio telescopes on the ground, 
would be able to further lower this limit to $\G\lesssim10^{-8}$. 

\end{abstract}


\begin{keywords}
gravitational lensing --- surveys --- cosmology: observations
\end{keywords}


\section{Introduction}

Superstrings of cosmic size \citep[introduced by ][]{Kib76}
are a generic prediction of a number of
string theory models \citep[see, e.g., ][and references therein]{Pol04,D+K05}. 
Given their macroscopic nature, they are in principle
detectable through astronomical observations. Therefore, they provide
a perhaps unique opportunity for direct empirical tests of the physics
of the very early Universe.

Considering strings whose only interactions are gravitational, all of
their effects are controlled by the global constant dimensionless
string tension~$\G$.  Current limits on this parameter are given
mainly by the properties of the cosmic microwave background (CMB) and
by studies of pulsar timing. As far as the former is concerned, cosmic
strings produce a smooth component in the CMB power spectrum and a
non-gaussian signature in the CMB anisotropy map
\citep[e.g.,][]{L+W05,J+S07}. As far as the latter is concerned,
strings produce a stochastic gravitational wave background,
detectable in the time series of pulsars
\citep[e.g.,][]{KTR94,D+V05}. 
Current limits are $\G \lesssim 3 \times 10^{-7}$ from
the CMB~\citep{PWW06}, and perhaps one or two orders of magnitude more stringent from
pulsar timing depending on the details of the statistical analysis and
the assumed string loop size distribution. An up-to-date review of current
observational limits is given by~\citet{Pol07}.

The idea of detecting cosmic strings by observing their gravitational
lensing effect dates back to~\citet{Vil84}. Briefly, cosmic
strings produce a conical space time resulting in a very clear strong
lensing signature, i.e.\ they produce identical (neither parity-flipped,
nor magnified or sheared) offset replica images of background
objects, separated by an angle proportional to the string tension.
Thus, strong gravitational lensing provides an opportunity for the
{\it direct} detection of cosmic strings, and even a single detected
event would provide a measurement of the string tension, independent
on the overall demographics of cosmic strings. A number of past
studies have identified cosmic string lens candidates in optical
surveys, but unfortunately none so far has withstood the test of
higher resolution imaging \citep{AHP06,Saz++07}.

In this paper we use current theoretical knowledge about the abundance
of cosmic strings to predict the number of lensing events as a
function of~$\G$ for a realistic set of current and future imaging
surveys. Our calculations show that, for sufficiently high angular
resolution and sufficiently high (yet currently allowable) string
tension, imaging surveys will either be able to detect cosmic string
lenses or to at least set interesting limits on the dimensionless
string tension parameter. For simplicity, we restrict our analysis to
long strings, i.e. strings that are the size of the cosmic volume, and
in particular straight with respect to the typical image separation,
neglecting the contribution from string loops. \citep[For a discussion of
the lens statistics in future radio surveys from the loop population
we refer the reader to the recent paper by ][]{MWK07}
This assumption simplifies significantly the treatment and,
since they do not depend on the detailed
topology of the string network, nor on the timescales for
gravitational decay, makes our predictions quite robust.


\section{Optical depth}
\label{sect:theory}

Let us consider a long string lens at angular diameter distance
$\Dd(\zd)$ from the observer, and a source at a distance $\Ds(\zs)$,
and let us denote the angular diameter distance between the string and
the source by $\Dds(\zd,\zs)$.  For a flat universe dominated by
matter and dark energy we have that:

\bea\label{r}
\Dd(\zd)=\frac{c}{H_0(1+\zd)}
         \int_0^{\zd}\frac{dz}{\sqrt{\Om(1+z)^3+\Ol}}
         \nonumber\\
\Dds(\zd;\zs)=\frac{c}{H_0(1+\zs)}
              \int_{\zd}^{\zs}\frac{dz}{\sqrt{\Om(1+z)^3+\Ol}},
\eea

\noindent
(and for $\Ds$ the same formula as for $\Dd$ but with exchanged
subscripts).  The source is lensed into a double image if it lies
behind the lens and within a strip of width

\be\label{dphi}
  d\beta_1=8\pi\Gfrac\frac{\Dds}{\Ds}|\sin i| 
\ee

\noindent
centered on the string, where $i$ is the angle between the tangent to
the string direction and the optical axis~\cite{Vil84}. This
$d\beta_1$ is then the cross-section per unit apparent length of
string, and we use the symbol $\beta$ to denote positions in the
source plane as usual. In order to find the lensing cross-section for
a source at a distance $\Ds$ lensed by a string at a distance $\Dd$ we
need to know how many radians $d\beta_2$ of long string are present at
$\Dd$.  

The physical length of long string lying in a shell of radius $\Dd(z)$
and depth $dz$, is given by the string mass in this shell $dM=\rho_{\rm str}
dV(z)$ divided by the string tension $\mu$, and so
$d\beta_2=\frac{dM}{\mu \Dd}|\sin i|$. Averaging over the inclination
angle $i$ ($\langle \sin^2 i \rangle=1/2$) the lensing cross-section
at a distance $\Dd(z)$ is given by
\be\label{int}
d\sigma= d\beta_1 \, d\beta_2 = \frac{dM}{\mu \Dd}\, 4\pi\Gfrac\frac{\Dds}{\Ds}.
\ee
In order to calculate the element of string mass $dM$, we assume that
the mass density in long strings follows the density of ordinary (dark
and baryonic) matter: {$\rho_{\rm str}=60\rho_{\rm m}\G$}~\citep{Pol04}. 
There is perhaps a factor of two
uncertainty on the prefactor in this expression. For clarity of
discussion, we choose to assert this formula and keep it in mind when
interpreting our bounds on $\G$, rather than introduce an exact
degeneracy between the string tension and the total string mass
density.

Writing $\rho_{\rm m}=\Om\frac{3H_0^2}{8\pi G}$ for the co-moving matter density,
and multiplying by the comoving volume element,
one gets straightforwardly for equation~(\ref{int}) that
\begin{align}\label{integ}
d\sigma &= 4\pi \, 90\Gfrac \, \frac{\Om H_0}{c} \, \nonumber \\ 
          & \frac{(1+\zd)^2}{\sqrt{\Om(1+\zd)^3+\Ol}} \, \frac{\Dd \Dds}{\Ds}\,d\zd.
\end{align}
Assuming that the cross-section overlaps are negligible with respect
to the total cross-section (we will see below that this
is the case), $\sigma(\zs)$ for the string lensing of a source
at a redshift $\zs$ is given by the integral of equation~(\ref{integ})
over all the possible string redshifts~$\zd$ such that the image
separation is observable~\citep[e.g.\ ][]{Sch06}:
\be\label{eq:totalXsection}
\sigma = \int_0^{\zs} \Theta(Z - \zd) \frac{d\sigma}{d\zd}\;d\zd.
\ee
We have here approximated the detection function as being the
Heaviside ``step" function, asserting that if the image separation is
greater than the instrument angular resolution~$\theta$ then it is
observable. This would be the case for point-like sources -- we return
to the issue of image detectability in section~\ref{sect:predict}
below. The limiting redshift~$Z$ in equation~\ref{eq:totalXsection} is that below
which the image separation is observable, i.e. such that the image
separation
\be\label{imlimit}
16\Gfrac\frac{\Dds(Z,\zs)}{\Ds}  = \theta
\ee
where we have again averaged over the inclination angle: $\langle| \sin
i |\rangle = 2/\pi$. If the string is too close to the source, the
image separation becomes too small to resolve -- but the fact that the
image separation is independent of the impact parameter means that
cosmic strings act as lenses no matter how close they are to the
observer, and this independence allows us to transform the angular
resolution limit of the observation into a limiting redshift for
observable strings.  The optical depth for strong lensing 
($\tau = \sigma/4\pi$) by long
strings is then
\begin{align}
\label{od}
&\tau(\zs,\theta,\G) = 90\Gfrac \, \frac{\Om H_0}{c} \, \nonumber \\ 
                     & \int_0^{\zs}  
                       \frac{(1+\zd)^2}{\sqrt{\Om(1+\zd)^3+\Ol}} \, 
                       \frac{\Dd \Dds}{\Ds}\,\Theta(Z-\zd)\,d\zd,&
\end{align}
which is easily computed numerically.  

\begin{figure*}
\begin{minipage}[t]{0.46\linewidth}
\centering\includegraphics[width=\linewidth]{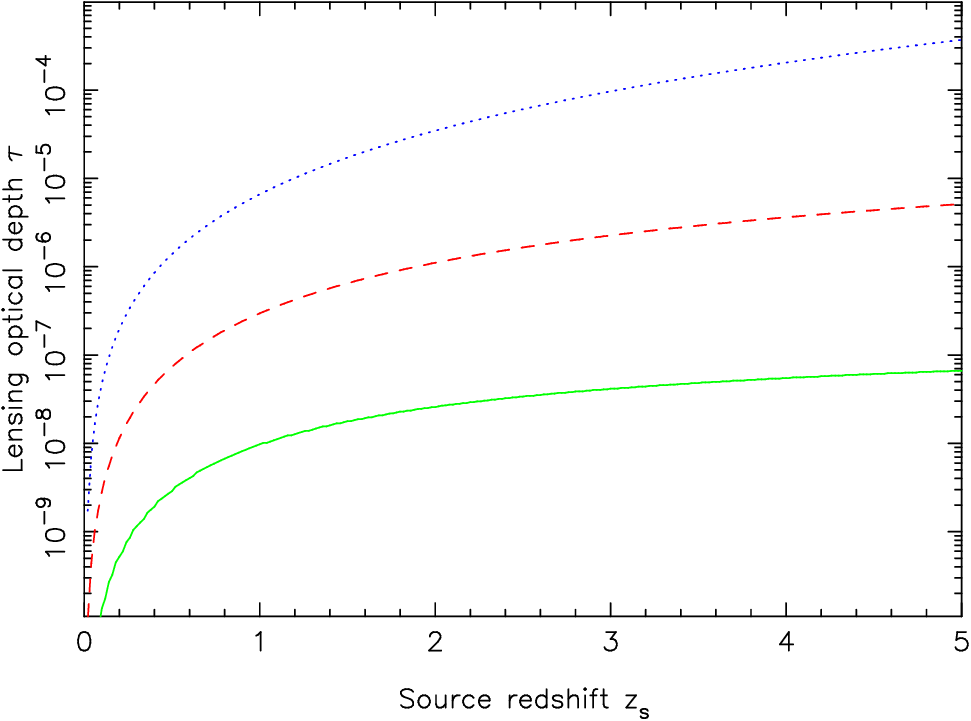} 
\end{minipage} \hfill
\begin{minipage}[t]{0.46\linewidth}
\centering\includegraphics[width=\linewidth]{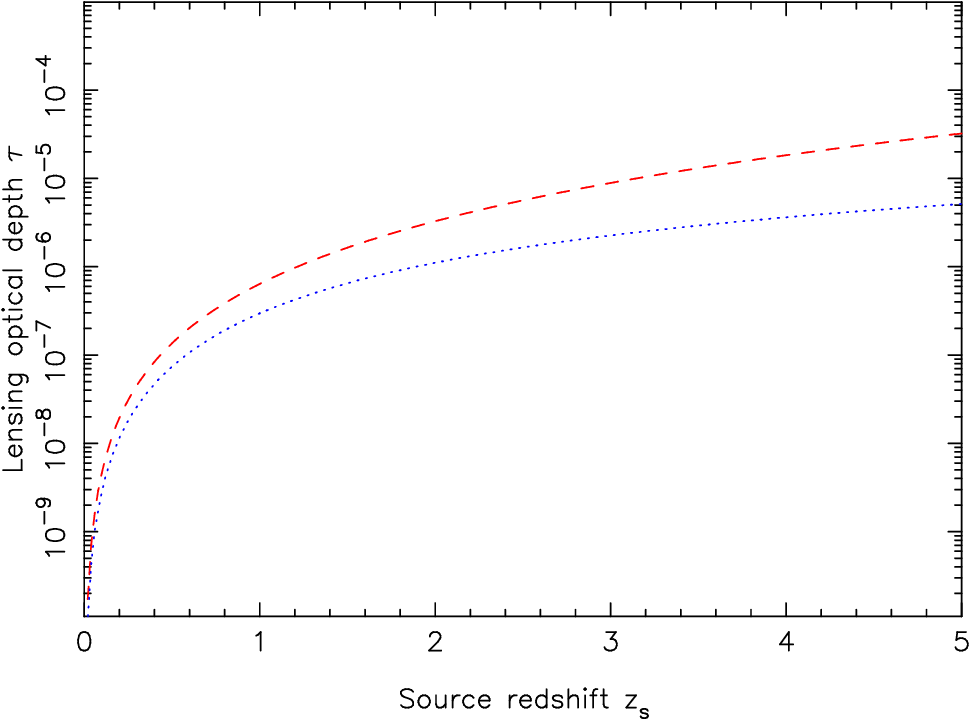} 
\end{minipage}
\caption{The optical depth $\tau$ as a function of source redshift.  
Left: with the angular resolution $\theta$ fixed at $0\farcs14$, we plot 
$\tau(\zs)$ for values of 
    $\G=10^{-6}$ (dotted), 
       $10^{-7}$ (dashed) and
$5\times10^{-8}$ (solid). 
Right: with $\G$ fixed at $10^{-7}$, we plot $\tau(\zs)$ for values of 
$\theta=0\farcs028$ (dashed), and
       $0\farcs14$ (dotted). The optical depth for image separations of 0.7
arcsec is negligible and does not appear on this plot.}\label{fig:tau}
\end{figure*}

From equation~(\ref{imlimit}) we can see that for each angular
resolution we consider, there is a minimum string tension below which
we would never be able to detect any lensing events (even with the
maximally-effective $\zd=0$ string position), and the optical depth
would be zero.  This ``angular resolution bound'' is $\G = (\theta /
16 {\rm rad}) \approx 3\times10^{-7} (\theta/{\rm arcsec})$. From
this we can already see that the limits from high resolution
(10-100~mas) imaging surveys will be interesting.  The optical depth 
$\tau$~is a
function of the source redshift $\zs$, of the string tension $\G$ and
of the experiment resolution angle $\theta$.  Figure~\ref{fig:tau}
shows the optical depth as a function of $\zs$ for different values of
$\G$ for a fixed $\theta$ (left panel), and for different values of
$\theta$ for a fixed $\G$ (right panel).
We assume $\Om = 0.3$, $\Ol = 0.7$ and $H_0 = 70 \,{\rm km s}^{-1} {\rm Mpc}^{-1}$
here and throughout this paper. 

The assumption behind equation (\ref{od}) is that the overlaps given by the
intersections of the long string projections on the sky are negligible with
respect to the total cross-section. We now show that this is a reasonable
assumption. Let us suppose we have~$n$ such intersections; then the
cross-section due to string intersections must be smaller than $\sim
n(8\pi\G)^2$. This is true if we require this quantity to be much smaller than
$\sigma$, or that $\tau \gg 16\pi n (\G)^2$. From Figure~\ref{fig:tau} we can
see that for a realistic source redshift values ($\zs>0.5$), 
the optical depth~$\tau$ is
of the order of~$\G$. Therefore the requirement is $n \ll \left(\G\right)^{-1}$. 

Recalling that the range of observation is well inside the Hubble horizon and
the fact that only a few dozen long strings per Hubble volume are expected
\citep[see, e.g., ][]{Pol04}, the only way to have $n\sim10^6$ or bigger is
through overlap of the cross-section of a string with itself. Furthermore, once
inside the horizon, the long strings tend to straighten through gravitational
waves emission and loop breaking \citep{P+R07}. As a consequence,
one do not expect a large amount of overlap for the range of redshift into
consideration.


\section{Expected number of observed string lensing events}
\label{sect:predict}

The expected number of lensing events for a given experiment is the
product $\Omega_{\rm s} \, N(\G,\theta),$ where $\Omega_{\rm s}$ is
the solid angle subtended by the survey in square degrees, and
$N(\G,\theta)$ is the expected number density of lensing events (in units of
deg$^{-2}$). $N$~in turn is given by the integral of the
optical depth over the source redshift distribution
$\frac{dN_{\rm s}}{d\zs}$, which, purely 
for concreteness, we take to be a Gaussian
peaked at redshift $\zs=1.5$ with width 0.4 normalized to 100 galaxies
per square arcmin, appropriate for optical surveys reaching a limiting
magnitude $I\sim 26$ mag \citep[e.g.][]{Ben++04}.  
This is a
conservative limit for space based surveys, which can reach
significantly higher density of (unresolved) galaxies. In any case, this
normalization factor is degenerate with the area of a survey, and
therefore one can easily generalize our results, trading area for
number of background galaxies.
\be 
N(\G,\theta)=
3.6\times10^{5}  \int_{0}^{\infty}
e^{-\frac{1}{2}\left(\frac{\zs-1.5}{0.4}\right)^2}\tau(\zs,\theta,\G)\,d\zs.
\ee

\begin{figure}
\begin{center}
\includegraphics[width=0.9\linewidth]{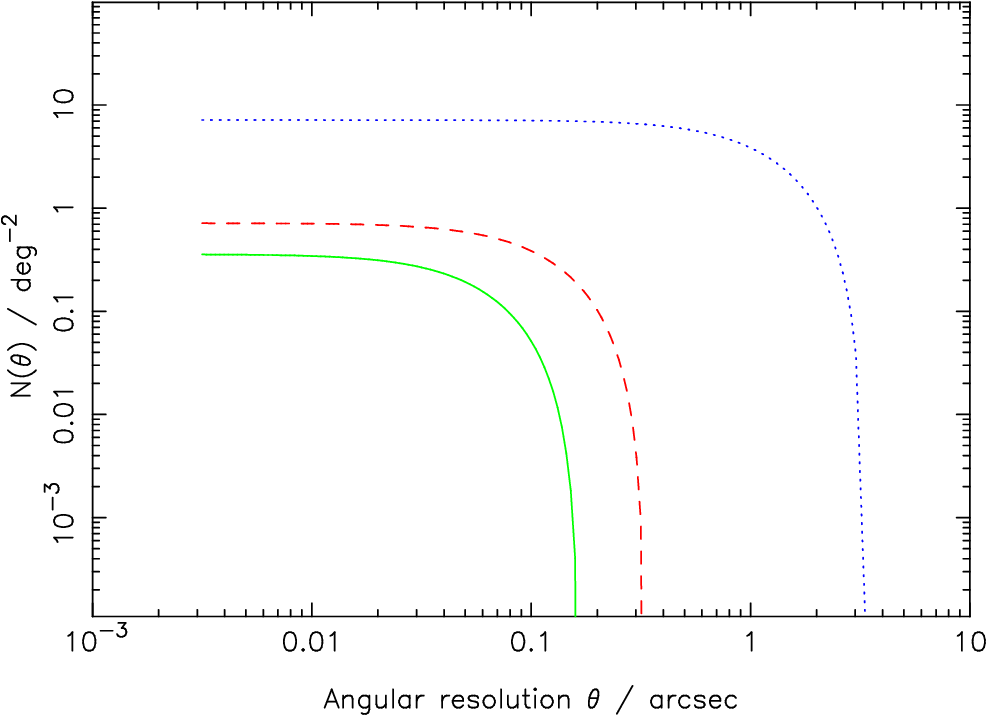} 
\caption{Predicted number density of lensing events $N$ per square degree as a function of the instrument angular resolution $\theta$/arcsec,
for the fixed values of 
    $\G=10^{-6}$ (dashed), 
       $10^{-7}$ (dotted) and
$5\times10^{-8}$ (solid).}\label{fig:bound}
\end{center}
\end{figure}

Figure~\ref{fig:bound} gives $N(\G,\theta)$ as a function of $\theta$
for fixed values of $\G=10^{-6},10^{-7}$ and $5\times10^{-8}$, assuming
a survey depth equivalent to recovering all 100 galaxies per square
arcmin. In practice this may not be possible at the lowest angular
resolution due to beam dilution, but as the steepness of
Figure~\ref{fig:bound} implies, it is the angular resolution of the
individual lens image pairs that is critical to the detection of the
lensing events.  In the next section we discuss the observational complications
associated with detecting lensing by cosmic strings in practice.

\section{Observational Issues}
\label{sect:obs}

We now turn briefly to the issue of practical lens system
detectability.   For a given string tension, and so image separation,
the maximal detectability will be achieved with the sources as compact
as possible, allowing them to be individually detected and flagged at
the object catalogue level. However, if the angular resolution is such
that the sources are well-resolved, then the possibility of detecting
the lensing effect using sharp edge-detection algorithms is opened up.
If the sources are resolved, morphological and colour information can be
used to verify the cosmic string lensing hypothesis. 

This somewhat idealistic case is altered if the string is moving
relativistically, if the string is not straight on the length scale of
the image separation in the plane of the string, or if the image is
``cut'' by the string so that one image is of only part of its source.
The detection of strings may  also be affected by the imaging survey
``footprint'' -- thus far we have effectively assumed that the sources
are observed one by one. We briefly discuss these possibilities in the
rest of this section.


\subsection{Relativistic strings}
\label{sect:obs:moving}

The relativistic movement of a string can significantly alter its lensing image
separation and cause an apparent redshift between the two images 
\citep{Vil86,S+T05}.
%
%
%
%
%
Both effects arise from Lorentz-transforming from the frame of
the string, in which the lensing occurs, into the frame of the observer and
source. A string with a velocity component parallel to the line of 
sight~$v_\parallel$
will have cross-section modified via
%
\be\label{eq:dbeta1prime}
  d\beta_1^\prime=\frac{\gamma d\beta_1}{(1-v_\parallel/c)},
\ee
which tends to infinity for highly relativistic strings!
A full statistical
treatment of the optical depth due to a network of fast-moving strings is beyond
the scope of this paper, but the above considerations remind us of two things:
first, that moving strings may be detectable by their lensing effects as well,
and second that any bound on~$\G$ is dependent on the assumptions of the string
network kinematics. In subsequent sections we assume the non-relativistic limit,
but note that current
estimates based on numerical simulations give $(v/c)^2 \sim 0.35$
(i.e.\ $(v_\parallel/c)^2 \sim 0.2$ and $\gamma \sim 1.5$) at the present time 
\citep{Pol07}, suggesting that our cross-sections are rather conservative.
In any case, relativistic corrections will have to be taken into 
account in future studies in order
to derive a precise limit for a specific survey.

A relativistically moving string with significant velocity in the plane of the
sky ($v_\perp$) will
also produce a redshift between the two sources given by 
%
\be\label{eq:motionredshift}
  \Delta z = \frac{\gamma v_\perp d\beta_1}{c},
\ee
where~$d\beta_1$ is measured in radians.
This redshift perturbation would cause an apparent colour difference
between the two images, warning us against the use of overly-restrictive cuts 
when
filtering catalogues. However, unless the string is highly relativistic
this redshift will be too small to be the dominant uncertainty in the
relative image colour.  

Bends in the string on the scale of the source size, and the ``cutting''
of sources whose image cross the line of the string,   produce
geometrical and even brightness distortions. The presence of many string
lensing events in a small region of space is really the robust signature
of the phenomenon \citep{H+V03,O+T05}, but quantifying the 
identifiability of complex lensing events with detailed image
simulations is possible. We describe an initial foray into this matter
in the next subsection.


\subsection{Practical image pair detection}
\label{sect:obs:pairs}

%
%
%
%

The lensing signal we are looking for is an overdensity of close pairs of
(resolved) galaxy images, aligned such that a string could pass through each
one. There are three  possible types of false positive for the individual
candidate lens systems: physical close galaxy pairs, line of sight projections,
and non-string gravitational lenses.  The latter do not concern us for three
reasons. First, the characteristic patterns of curved and distorted images in
conventional galaxy-galaxy strong  lenses \citep[see e.g.\ ][for
examples]{Bol++06,Mou++07} are quite different from the translated and truncated
images expected for a string lens \citep[see e.g.\ ][for examples]{Saz++07}.
Second, in optical images of galaxy-galaxy strong  lenses the lens galaxies 
are usually brighter than the images of the source, confirming the nature of the
lens. Third, at a mean density of $\sim 10$ per square degree \citep{MBS05},
ordinary galaxy-scale strong lenses are sufficiently rare that one does not
expect to find more than one in a field of view of a few square arcminutes
\citep[although see ][for a notable counter-example]{Fas++06}.

The first two types of false positive -- projected and physical close pairs of
galaxies -- can be rejected using high quality follow-up data \citep[as in the
case of CSL-1,][]{Saz++07}. Alternatively, during a survey they can be dealt with
statistically via the observed galaxy two-point angular correlation function, a
topic of active research by our group  (Morganson et al., in preparation).  
In simulated
HST F606W-band images with limiting magnitude 28 (comparable to the GOODS
survey, for example, and assumed to contain background sources at a number
density of 240~arcmin$^{-2}$), we find (on average) 4~faint galaxy pairs
separated by~$0\farcs5$ in a 1~arcsec by 5~arcmin image  strip.  We can compare
this result to images that have been lensed by simulated strings.  For
illustration we consider a fiducial non-relativistic string  of tension $\G =
1.2 \times 10^{-7}$ at $\zd = 0.5$ lying in the plane of the sky, and a
population of compact, faint sources at $\zs = 1.5$. This set-up leads to a
convenient source plane  cross-section  of width $d\beta_1 = 0\farcs5$.  We
generated mock background scenes, and then overlaid the string and recomputed
the surface brightness where lensing occurs, then convolved the images with a
mock HST PSF (FWHM $\approx 0\farcs14$) and added background and source Poisson
noise  consistent with the counts expected in an HST image of limiting
magnitude~28.

In these simulated images we are able to detect an average of 14~faint  galaxy
pairs separated by by~$0\farcs5$ in the same 1~arcsec  by 5~arcmin  strip
(including 10 that are multiple images caused by the string).   This number is
consistent with that calculated by \citet{Saz++07}. This simple observation of
extra pairs is robust to moderate changes in color, magnitude and morphology
caused by non-ideal strings, and would, in principle, require only one deep HST
image in the right patch of sky -- the simulated detection described above has a
significance of more than 3-sigma. Moreover, the condition that the image pairs
be aligned perpendicular to  the putative string, and the requirement that all
background objects be lensed if they are positioned behind the string,  will
allow the rejection of the majority of false positives.


\subsection{Survey geometry}
\label{sect:obs:surveygeom}

The optical depth is most relevant to a source-oriented survey, where
all the sources in the sky  are observed. The deduced lens number
density calculated above is therefore  an \emph{average event rate}. The
usual cosmological assumptions of isotropy and homogeneity do not
necessarily apply to cosmic string lensing events.  The possibility of
having only a few long strings is favored by most models: the area
around these strings could contain many detectable lensing events while
the remaining (vast) expanse of sky would contain none.  In practice,
observing costs limit us to imaging surveys of a fixed area of the sky,
which will include blank sky as well as sources and will be composed of
some number of finite contiguous fields.  The survey geometry is
therefore rather important when the survey area is small, as we now
show  with a simple geometrical example. 

Let us assume that we have a survey of total solid angle~$A$ which is composed
of a number of {\em randomly distributed} square fields each of side~$2r$
radians, and area $a = 4r^2$ steradians.  Let us also assume that we have a 
single straight string of angular length~$L$ which will be detected if it is 
%
%
%
within~$r$ of the center of one of our fields.
The probability 
any given field will detect our string is roughly the fraction of sky that is 
within~$r$ of the string:
\begin{align}\label{field}
 P_{\rm
 field}&=\frac{1}{4\pi}\int_{-r}^{r}\int_{-r}^{L+r}\cos{\theta_1}d\theta_1
 d\theta_2\nonumber\\
              &=\frac{2(L+2r)\sin r}{4\pi} \approx \frac{L\sqrt{a}}{4\pi}.
\end{align}
The probability of the survey containing a string in at least one of 
the $N = A/a$ fields is then
\begin{align}\label{survey}
 P_{\rm survey} &= 1 - (1 - P_{\rm field})^{N}\nonumber\\
                 &\approx 1 - {\rm e}^{-\frac{N L \sqrt{a}}{4\pi}}\nonumber\\
                &\approx 1 - {\rm e}^{-\frac{L \sqrt{N A}}{41250 {\rm deg}^2}}
\end{align}
This holds providing our many strings have cumulative length $L$ sufficiently
long that the individual strings are much longer than $\sqrt{a}$, so that their overlap
region is not large (Section~\ref{sect:theory}). Also, it is clear that the
random distribution of the fields is an important assumption in 
this simple calculation. Non-randomly chosen fields would have to be treated
with more care.

As a practical example, we note that our assumption that  $\rho_{\rm
str}=60\rho_{\rm m}\G$ is equivalent to having 2250 degrees of string at
$z < 0.5$, independent of the string tension  (recalling that
simulations of string interactions indicate that the actual string 
density is within a factor two of this estimate).  The independence
arises because the string density is proportional to the tension, while
the string length is just proportional to  the density divided by the
tension.

Assuming this string density then,  a 5 square-degree survey divided
into 1800 randomly-distributed small fields  (each of e.g.\ $10$
arcmin$^2$, approximately the field of view of the Advanced Camera for
Surveys aboard the Hubble Space Telescope) will contain a string with
more than 99\% probability. However, with the same density and survey
area but  a contiguous survey geometry, the  string would only fall in
the survey region with 11\%  probability, and we would not be able to 
confidently place a limit on $\mu$.

Conversely, surveys of more than a few square degrees will reliably 
contain strings even without many divisions.   We find that, in order
to  achieve a 95\% probability of including a long string, surveys of
area 0.5, 5, and 1000 square degrees must be divided into 6000 fields,
600 fields and 3 fields respectively (assuming the string density of
section~\ref{sect:theory}). Total survey areas greater than these  are
large enough to be guaranteed a string crossing, regardless of their
geometry.  For the rest of this paper we assume that the imaging survey
in question is composed of randomly distributed fields that are 
sufficiently small to make the average  lensing event rate given by the
theoretical optical depth appropriate for interpreting the observations.


\section{Bounds on $\G$}
\label{sect:bound}

If a single detection of string lensing was made, then the optical
depth would not be zero, and a lower bound could be put on the string
tension $\G$.  This bound would be given by the angular resolution of
the experiment, as described at the end of
section~\ref{sect:theory}. It is a lower bound since the maximum image
separation occurs when the string is right in front of the observer --
moving the string to higher redshift would require a larger tension to
keep the image separation as large as observed. If somehow the string
redshift can be constrained, statistically or via the presence of
foreground non-lensed objects, then the image separation provides
direct measurement of the string tension.


On the other hand, if nothing is observed we cannot state that the
upper bound on $\G$ is given by this ``angular resolution bound''
argument.  We expect the upper bound to be bigger than
$\frac{\theta}{16}$, with the exact amount by which it is bigger
dependent on the observed sky area $\Omega_{\rm s}$, and the source
density distribution $\frac{dN_{\rm s}}{d\zs}.$ Treating string lensing
events as rare, and so using the Poisson distribution for the
probability of their occurrence given the predicted mean rate
(previous section), then if no string lensing detections are made, we
can state at $95\%$ confidence that the upper bound on $\G$ is given
by the root of $N(\G,\theta)=\frac{2.996}{\Omega_s}$~\citep{Geh86}. 


\begin{table*}
\begin{tabular}{c|cccc}
Angular     & & Survey area (deg$^2$)&& \\
Resolution	& $0.5$                &  5	                 &  1000         &  20,000\\
\hline\hline
0$ \farcs 028 $ & $ < 8.4 \times 10^{-7} $ & $ < 9.1 \times 10^{-8} $ & $ < 1.1 \times 10^{-8} $ & $ < 9.0 \times 10^{-9} $ \\
0$ \farcs 14 $  & $ < 8.6 \times 10^{-7} $ & $ < 1.5 \times 10^{-7} $ & $ < 4.7 \times 10^{-8} $ & $ < 4.3 \times 10^{-8} $ \\
0$ \farcs 7 $   & $ < 1.1 \times 10^{-6} $ & $ < 4.0 \times 10^{-7} $ & $ < 2.2 \times 10^{-7} $ & $ < 2.1 \times 10^{-7} $ \\
\hline
\end{tabular}
\caption{Inferred upper limits (95\%) on the string tension 
corresponding to no detection, 
for a variety of combinations of instrumental resolution and survey area. 
\label{tab:surveys}}
\end{table*}

\begin{figure}
\begin{center}
\includegraphics[width=0.95\linewidth]{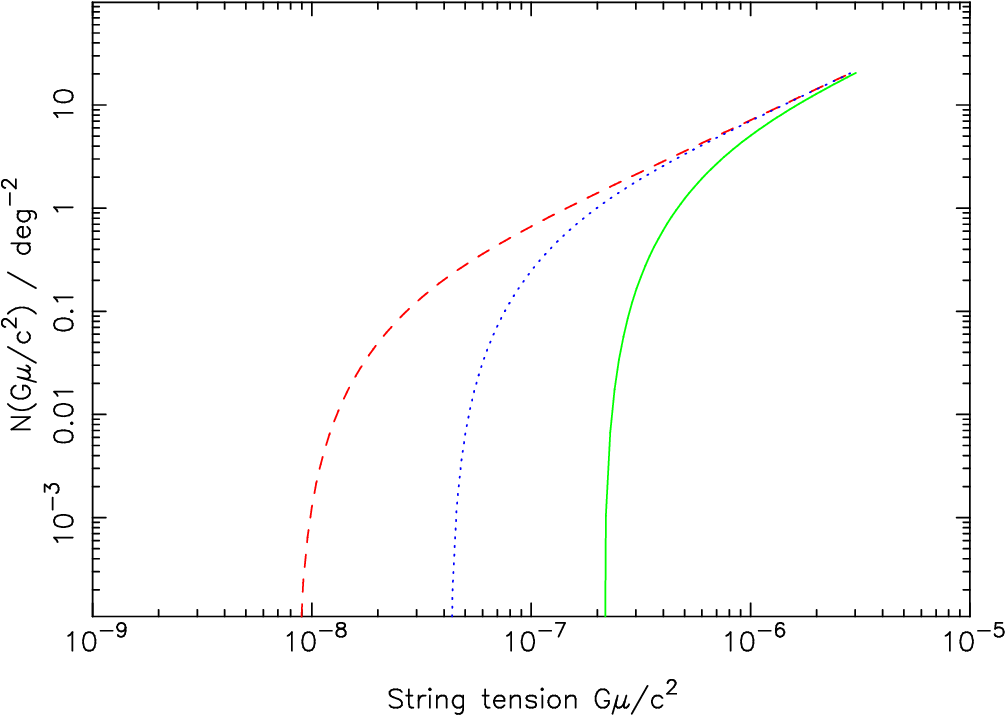} 
\caption{The predicted number density of string lensing events $N$
per square degree as a function of $\G$, for three values of the angular resolution
$\theta=0\farcs028$ (dashed), 
       $0\farcs14$ (dotted) and 
       $0\farcs7$ (solid).}\label{fig:numbers}
\end{center}
\end{figure}

We now place our simple theoretical calculation in observational
context, building on the discussion in section~\ref{sect:obs}. 
We anticipate requiring surveys covering large areas of sky,
and observing high surface densities of faint galaxies at high angular
resolution, but aim to show the link between what is possible now and
what may be possible in the future, wide-field era. We investigate
three representative angular resolutions, typical of ground-based
optical imaging 
($0\farcs7$), of space-based optical/infra-red imaging 
($0\farcs14$), and of planned radio surveys 
($0\farcs028$).  (This last resolution may also be achievable with next
generation adaptive optics on large optical telescopes,
although the area surveyed will be necessarily smaller.)  We have
purposely tried to be conservative in these angular resolutions. For
simplicity we do not account for the different survey depths. In most
cases the number of background sources will be irrelevant beyond some
threshold, as will be the area of the surveys, since the factor
dominating the sensitivity is the angular resolution. Figure~\ref{fig:numbers}
shows the expected number density of string lensing events, 
as a function of $\G$, for our three 
fiducial survey resolutions.

Table~\ref{tab:surveys} shows upper limits on $\G$ in the event of a
null survey result for surveys of various solid angles.
square degrees.  The first two choices of sky area (0.5 and 5 square degrees)
represent high resolution surveys within current capabilities 
(e.g., the HST-ACS archive, Marshall et al. 2007, in prep). 
The larger survey areas, 1000 and 20000,
square degrees, represent reasonable expectations for the
next generation of space and ground based surveys, 
respectively~\citep[see e.g.][]{SNAP,LSST}.

Let us illustrate this table a little.  For example, for a fixed resolution
of $\theta=0\farcs14$, choosing $\Omega_{\rm s} = 5$ square degrees
gives $\G \leq 1.5\times 10^{-7}$.  
This limit is about a factor of two lower than the most recent
$95\%$ confidence bound from studies of the cosmic microwave
background~\citep{PWW06,S+S06,J+S07}, and is competitive with the pulsar
timing studies by virtue of having different model dependencies (the reader
is referred to \citet{Pol07} for a critical review of current
observational bounds on~$\G$).  

Note that, because of the steep shape of the curve near this value
(Figure~\ref{fig:numbers}), observing a larger area of sky without 
(long) string lensing observations would not give a proportionate decrease 
in the bound on$~\G$.  Even though this may appear surprising at first, this is a
simple consequence of the fact that the main limitation in fixing the 
bound comes from the angular resolution.

In other words, for any given resolution there is a maximum area that is worth
exploring for strings, the exact value of which  will depend on the survey
geometry, the string correlation length and the image depth (i.e.\ on the number 
of available background source galaxies). For example, at ground-based optical
image resolution ($0\farcs7$), a survey of $\sim30$~square degrees is sufficient to reach 
saturation -- and even then the resulting upper
limit on the string tension is above that available from a few square degrees
of space-based resolution optical imaging.
At the highest resolution considered here, this maximum area is
correspondingly higher; however, we can see that the difference between 1000 square
degrees and half the sky is small (a mere 20\% improvement in the bound
on~$\G$). To make progress then, one needs better 
resolution first, and then more area.  

This conclusion has two practical consequences. On 
the one hand, data already exist
that can probe the $\G\sim10^{-7}$ regime, with 
a suitably robust algorithm (Morganson et al.,  in preparation). 
On the other hand, we have not yet reached the maximum useful 
area for space-based optical string lens surveys: the turnover in 
Figure~\ref{fig:numbers} for $0\farcs14$ resolution is at a few hundred square
degrees. Moving to even higher resolution at the same scale of survey will 
result in a
gain of a further factor of~5 or so, ultimately 
providing sensitivity to string tensions 
of~$\G\sim10^{-8}$. 

\section{Conclusions}
\label{sect:concl}

We have used a robust prediction of theoretical work on cosmic string
networks to predict, in a straightforward way, the expected number of strong
gravitational lensing events visible in astronomical imaging surveys of varying
angular resolution and sky coverage. From our simple analysis we draw the
following conclusions:

\begin{enumerate}

\item Present-day high-resolution imaging surveys are capable of probing the
putative cosmic string tension parameter to a factor of two lower than the
current CMB limit, making lensing both competitive with, and complementary to,
the pulsar timing methods.
As has been noted before, in the event of a detection, gravitational lensing
would provide a direct measurement of the tension, and perhaps the velocity, of
this string. 

\item The main practical considerations in detecting string lensing events are
two-fold: firstly, the expected faint pairs of images must first be carefully 
deblended and then understood in the context of neighbouring events; secondly,
the survey geometry should be such that the fields are large enough to contain
the characteristic multiple neighbouring events, but sparsely distributed to
ensure that the global, not local, lensing rate is being probed.

\item The upper bound on the tension, from the failure to detect a single string
lensing event in a given survey, is principally determined by the available
angular resolution. Relatively little is gained from studying an area of sky
greater than some critical value: for typical ground-based optical image 
resolution, this critical survey size is a few tens of square degrees.  
Likewise, string tensions of $\G\sim10^{-8}$ should be able to be investigated
in future surveys with image resolutions of 10-100~milliarcsec
covering a few hundred square degrees.

\end{enumerate}


\section*{Acknowledgments} 

We thank Joe Polchinski for inspiring this work with a Blackboard Lunch Talk at
the Kavli Institute for Theoretical Physics (KITP) and for numerous useful
suggestions. We thank David Hogg and Roger Blandford for useful discussions. 
We
are grateful for the comments of the anonymous referee, that led to a  clearer
explanation of our work.  
AG, PJM, and TT would
like to thank the KITP and its staff for the warm hospitality during the
programs ``Applications of gravitational lensing'' and ``String phenomenology''
when a significant part of the work presented here was carried out (KITP is
supported by NSF under grant No.\ PHY99-07949).  PJM received support from the
TABASGO foundation in the form of a research fellowship.  TT acknowledges
support from the NSF through CAREER award NSF-0642621, and from the Sloan
Foundation through a Sloan Research Fellowship. The work of FD was 
supported by Swiss National Funds and by the
NSF under grant No.\ PHY99-07949 (through KITP). 
This work  was supported in part
by the U.S.\ Department of Energy under contract  number
DE-AC02-76SF00515.                                 

\label{lastpage}




\bsp
\end{document}